\documentclass{amsart}
\usepackage{hyperref}

\newtheorem{theorem}{Theorem}[section]
\newtheorem{lemma}[theorem]{Lemma}
\newtheorem{corollary}[theorem]{Corollary}
\newtheorem{remark}[theorem]{Remark}
\newtheorem{hypo}[theorem]{Hypothesis {\bf H.}\hspace*{-0.6ex}}

\newcommand{\R}{{\mathbb R}}
\newcommand{\N}{{\mathbb N}}
\newcommand{\Z}{{\mathbb Z}}
\newcommand{\C}{{\mathbb C}}

\newcommand{\nn}{\nonumber}
\newcommand{\be}{\begin{equation}}
\newcommand{\ee}{\end{equation}}
\newcommand{\bea}{\begin{eqnarray}}
\newcommand{\eea}{\end{eqnarray}}
\newcommand{\ba}{\begin{array}}
\newcommand{\ea}{\end{array}}

\newcommand{\ti}{\tilde}

\newcommand{\spr}[2]{\langle #1 , #2 \rangle}

\newcommand{\lz}{\ell^2(\Z)}
\newcommand{\tl}{\mathrm{TL}}
\newcommand{\km}{\mathrm{KM}}

\numberwithin{equation}{section}

\begin{document}

\title[Equivalence of Lax Pairs for the Kac--van Moerbeke Hierarchy]{On the Equivalence of Different
Lax Pairs for the Kac--van Moerbeke Hierarchy}

\author[J. Michor]{Johanna Michor}
\address{Imperial College\\
180 Queen's Gate\\ London SW7 2BZ\\ and International Erwin Schr\"odinger
Institute for Mathematical Physics, Boltzmanngasse 9\\ 1090 Wien\\ Austria}
\email{\href{mailto:Johanna.Michor@esi.ac.at}{Johanna.Michor@esi.ac.at}}
\urladdr{\href{http://www.mat.univie.ac.at/~jmichor/}{http://www.mat.univie.ac.at/\~{}jmichor/}}

\author[G. Teschl]{Gerald Teschl}
\address{Faculty of Mathematics\\
Nordbergstrasse 15\\ 1090 Wien\\ Austria\\ and International Erwin Schr\"odinger
Institute for Mathematical Physics, Boltzmanngasse 9\\ 1090 Wien\\ Austria}
\email{\href{mailto:Gerald.Teschl@univie.ac.at}{Gerald.Teschl@univie.ac.at}}
\urladdr{\href{http://www.mat.univie.ac.at/~gerald/}{http://www.mat.univie.ac.at/\~{}gerald/}}

\thanks{Work supported by the Austrian Science Fund (FWF) under Grants No.\ Y330 and J2655.}


\begin{abstract}
We give a simple algebraic proof that the two different Lax pairs for the Kac--van Moerbeke
hierarchy, constructed from Jacobi respectively super-symmetric Dirac-type
difference operators, give rise to the same hierarchy of evolution equations.
As a byproduct we obtain some new recursions for computing these equations.
\end{abstract}

\maketitle

\section{Introduction}

There are two different Lax equations for the Kac--van Moerbeke equation:
The original one of Kac and van Moerbeke \cite{km} based on a Jacobi matrix with zero diagonal elements and its skew-symmetrized square and the second one based on super-symmetric
Dirac-type matrices. Both approaches can be generalized to give corresponding
hierarchies of evolution equations in the usual way and both reveal a close connection to the
Toda hierarchy. In fact, the first approach shows that the Kac--van Moerbeke hierarchy (KM hierarchy)
is contained in the Toda hierarchy by setting $b=0$ in the odd equations. The second
one relates both hierarchies via a B\"acklund transformation since the  Dirac-type
difference operator gives rise to two Jacobi operators by taking squares (respectively
factorizing positive Jacobi operators to obtain the other direction). Both ways of
introducing the KM hierarchy have its merits, however, tough it is {\em obvious}
that both produce the same hierarchy by looking at the first few equations, we
could not find a formal proof in the literature. The purpose of this short note is to give
a simple algebraic proof for this fact. As a byproduct we will also obtain some new
recursions for computing the equations in the KM hierarchy.

In Section~\ref{secth} we review the recursive construction of the Toda hierarchy
via Lax pairs involving Jacobi operators and obtain the KM hierarchy
by setting $b=0$ in the odd equations. In Section~\ref{seckm} we introduce
the KM hierarchy via Lax pairs involving Dirac-type difference operators.
In Section~\ref{secmain} we show that both constructions produce the same equations.
Finally, we recall how to identify Jacobi operators with $b=0$ in Section~\ref{app}.

\section{The Toda hierarchy}
\label{secth}

In this section we introduce the Toda hierarchy using the standard Lax formalism
following \cite{bght} (see also \cite{tjac}).

We will only consider bounded solutions and hence require

\begin{hypo} \label{habt}
Suppose $a(t)$, $b(t)$ satisfy
\[
a(t) \in \ell^{\infty}(\Z, \R), \qquad b(t) \in \ell^{\infty}(\Z, \R), \qquad
a(n,t) \neq 0, \qquad (n,t) \in \Z \times \R,
\]
and let $t \mapsto (a(t), b(t))$ be differentiable in 
$\ell^{\infty}(\Z) \oplus \ell^{\infty}(\Z)$.
\end{hypo}

\noindent
Associated with $a(t), b(t)$ is a Jacobi operator
\begin{equation} \label{defjac}
H(t) = a(t) S^+  + a^-(t) S^-  + b(t)
\end{equation}
in $\lz$, where $S^\pm f(n) = f^\pm(n)= f(n\pm1)$ are the usual shift operators and
$\lz$ denotes the Hilbert space of square summable (complex-valued) sequences
over $\Z$. Moreover, choose constants $c_0=1$, $c_j$, $1\le j \le r$, $c_{r+1}=0$, and set
\begin{align} 
\begin{split}
g_j(n,t) &= \sum_{\ell=0}^j c_{j-\ell} \spr{\delta_n}{H(t)^\ell \delta_n},\\ \label{todaghsp}
h_j(n,t) &= 2 a(n,t) \sum_{\ell=0}^j c_{j-\ell}  \spr{\delta_{n+1}}{H(t)^\ell
\delta_n} + c_{j+1}.
\end{split}
\end{align}
The sequences $g_j$, $h_j$ satisfy the recursion relations
\begin{align}
\nn g_0 = 1, \: h_0 &= c_1,\\ \nn
2g_{j+1} -h_j -h_j^- -2b g_j &= 0,\quad 0 \le j\le r,\\ \label{rectodah}
h_{j+1} -h_{j+1}^- - 2(a^2 g_j^+ -(a^-)^2 g_j^-) - b
(h_j -h_j^-) &= 0, \quad 0 \le j < r.
\end{align}
Introducing
\begin{equation}  \label{btgptdef}
P_{2r+2}(t) = -H(t)^{r+1} + \sum_{j=0}^r ( 2a(t) g_j(t) S^+ -h_j(t)) H(t)^{r-j} +
g_{r+1}(t),
\end{equation}
a straightforward computation shows that the Lax equation
\begin{equation} \label{laxp}
\frac{d}{dt} H(t) -[P_{2r+2}(t), H(t)]=0, \qquad t\in\R,
\end{equation}
is equivalent to
\begin{equation}\label{tlrabo}
\tl_r (a(t), b(t)) = \begin{pmatrix} \dot{a}(t) -a(t) \Big(g_{r+1}^+(t) -
g_{r+1}(t) \Big)\\ 
\dot{b}(t) - \Big(h_{r+1}(t) -h_{r+1}^-(t) \Big) \end{pmatrix} =0,
\end{equation}
where the dot denotes a derivative with respect to $t$.
Varying $r\in \N_0$ yields the Toda hierarchy
$\tl_r(a,b) =0$. The corresponding homogeneous quantities obtained
by taking all summation constants equal to zero, $c_\ell \equiv 0$, $\ell \in \N$, are denoted
by $\hat g_j$, $\hat h_j$, etc., resp.\ 
\begin{equation}
\widehat{\tl}_r (a, b) = \tl_r (a, b)\big|_{c_\ell\equiv0, 1\leq \ell \leq r}.
\end{equation}

Next we show that we can set $b \equiv 0$ in the odd equations of the Toda hierarchy.

\begin{lemma}
Let $b \equiv 0$. Then the homogeneous coefficients satisfy 
\[
\hat g_{2j+1}=\hat h_{2j}=0, \quad j \in \N_0. 
\]
\end{lemma}
\begin{proof}
We use induction on the recursion relations \eqref{rectodah}. The claim is true for $j=0$. If 
$\hat h_{2j}=0$ then $\hat g_{2j+1}=0$, and $\hat h_{2j}=0$ follows from the last 
equation in \eqref{rectodah}.
\end{proof}

In particular, if we choose $c_{2\ell}=0$ in $\tl_{2r+1}$, then we can set $b \equiv 0$
to obtain a hierarchy of evolution equations for $a$ alone. In fact, set
\begin{equation} \label{defGTL}
G_j = \hat g_{2j}, \quad K_j= \hat h_{2j+1},
\end{equation}
in this case. Then they satisfy the recursion
\begin{align}
\nn G_0 = 1, \quad K_0=2a^2,\\ \nn
2G_{j+1} -K_j -K_j^- &= 0,\quad 0 \le j\le r,\\ \label{reckmtl}
K_{j+1} -K_{j+1}^- - 2(a^2 G_j^+ -(a^-)^2 G_j^-) &= 0, \quad 0 \le j < r,
\end{align}
and $\tl_{2r+1}(a,0)=0$ is equivalent to the KM hierarchy defined as
\begin{equation}
\km_r(a) = \dot{a} - a (G_{r+1}^+ -G_{r+1}), \qquad r\in\N_0.
\end{equation}

\section{The Kac--van Moerbeke hierarchy as a modified Toda hierarchy}
\label{seckm}

In this section we review the construction of the KM hierarchy as
a modified Toda hierarchy. We refer to \cite{bght}, \cite{tjac} for further details.

Suppose $\rho(t)$ satisfies

\begin{hypo} \label{hrho}
Let
\begin{equation} 
\rho(t) \in\ell^\infty(\Z,\R), \quad \rho(n,t)\neq 0, \;  (n,t)\in\Z\times\R
\end{equation}
and  let $t \mapsto \rho(t)$ be differentiable in $\ell^\infty(\Z)$. 
\end{hypo}

Define the ``even'' and ``odd'' parts of $\rho(t)$ by
\begin{equation} \label{rhoeo}
\rho_e (n,t) =\rho(2n,t), \; \rho_o (n,t) =\rho (2n+1,t), \quad (n,t)
\in\Z\times \R,
\end{equation}
and consider the bounded operators (in $\ell^2 (\Z)$)
\begin{equation}
A(t) =\rho_o(t) S^+ + \rho_e(t), \; A(t)^* =\rho_o^-(t) S^- +\rho_e(t).
\end{equation}
In addition, we set
\begin{equation}
H_1(t) =A(t)^* A(t), \quad H_2 (t) =A(t) A(t)^*,
\end{equation}
with
\begin{equation}
H_k(t) =a_k (t) S^+ +a_k^- (t) S^- +b_k (t), \qquad k =1,2,
\end{equation}
and
\bea \label{defaot}
a_1(t) = \rho_e(t) \rho_o(t), &\qquad& b_1(t) = \rho_e(t)^2 +\rho_o^-(t)^2,
\\ \label{defatt}
a_2(t) = \rho_e^+(t) \rho_o(t), &\qquad& b_2(t) = \rho_e(t)^2 +\rho_o(t)^2.
\eea

Now we define operators $D(t)$, $Q_{2r+2}(t)$ in $\ell^2(\Z,\C^2)$ as follows,
\bea
D(t) &=& \left( \ba{cc} 0 & A(t)^* \\ A(t) & 0 \ea \right),\\ \label{defQtrpt}
Q_{2r+2}(t) &=& \left( \ba{cc} P_{1,2r+2}(t) & 0 \\ 0 & P_{2,2r+2}(t)
\ea \right), \quad r\in\N_0.
\eea
Here $P_{k,2r+2}(t)$, $k=1,2$ are defined as in (\ref{btgptdef}),
that is,
\begin{equation}
P_{k,2r+2} (t) = -H_k(t)^{r+1} +\sum_{j=0}^r ( 2 a_k(t)
g_{k,j}(t) S^+ - h_{k,j}(t)) H_k(t)^j +g_{k,r+1},\\
\end{equation}
$\{g_{k,j}(n,t)\}_{0\le j \le r}$, $\{h_{k,j}(n,t)\}_{0\le j \le r+1}$ are defined as in (\ref{todaghsp}).
Moreover, we choose the same integration constants in $P_{1,2r+2}(t)$ and $P_{2,2r+2}(t)$ (i.e.,
$c_{1,\ell}=c_{2,\ell} \equiv c_\ell, \: 1 \le \ell \le r$).

Analogous to equation (\ref{laxp}) one obtains that
\begin{equation} \label{laxkm}
\frac{d}{dt} D(t) - [Q_{2r+2}(t), D(t)] =0
\end{equation}
is equivalent to
\bea \nn
\km_r(\rho) &=& (\km_r (\rho)_e, \; \km_r(\rho)_o)\\
&=& \left( \ba{cc}
\dot{\rho}_e - \rho_e(g_{2,r+1} -g_{1,r+1}) \\
\dot{\rho}_o + \rho_o(g_{2,r+1} -g_{1,r+1}^+) \ea \right) =0.
\label{ulkmhie}
\eea
As in the Toda context (\ref{tlrabo}), varying $r\in\N_0$ yields the KM
hierarchy which we denote by
\begin{equation} \label{kmhie}
\km_r(\rho) =0, \quad r\in\N_0.
\end{equation}
The homogeneous $\km$ hierarchy is denoted by
\begin{equation}
\widehat \km_r(\rho) =  \km_r (\rho)\big|_{c_\ell\equiv0, 1\leq \ell \leq r}.
\end{equation}
One look at the transformations \eqref{defaot}, \eqref{defatt} verifies that the 
equations for $\rho_o$, $\rho_e$ are in fact
one equation for $\rho$. More explicitly, combining $g_{k,j}$, resp.\ $h_{k,j}$, into
one sequence
\begin{equation} \label{defGKM}
\ba{lcl} G_j(2n) &=& g_{1,j}(n)\\ G_j(2n+1) &=& g_{2,j}(n)\ea, \mbox{ resp.\ }
\ba{lcl} H_j(2n) &=& h_{1,j}(n)\\ H_j(2n+1) &=& h_{2,j}(n)\ea,
\end{equation}
we can rewrite (\ref{ulkmhie}) as
\begin{equation}
\km_r(\rho) = \dot{\rho} - \rho(G_{r+1}^+ -G_{r+1}).
\end{equation}
From (\ref{rectodah}) we see that $G_j$, $H_j$ satisfy the recursions
\begin{align}
\nn G_0 = 1, \: H_0 &= c_1,\\ \nn
2G_{j+1} -H_j -H_j^{--} -2(\rho^2 + (\rho^-)^2) G_j &= 0, \quad 0 \le j\le r,\\ \nn
H_{j+1} -H_{j+1}^{--} - 2((\rho\rho^+)^2 G_j^+ -(\rho^-\rho)^2 G_j^{--}) &\\ \label{reckm}
 - (\rho^2 + (\rho^-)^2) (H_j -H_j^{--}) &= 0, \quad 0 \le j < r.
\end{align}
The homogeneous quantities are denoted by $\hat G_j$, $\hat H_j$, etc., as before. 

As a simple consequence of (\ref{laxkm}) we have
\begin{equation}
\frac{d}{dt} D(t)^2 - [Q_{2r+2}(t), D(t)^2] =0
\end{equation}
and observing
\begin{equation}
D(t)^2 = \left(\ba{cc} H_1(t) & 0\\ 0 & H_2(t) \ea\right)
\end{equation}
yields the implication
\begin{equation} \label{kmimpltl}
\km_r(\rho) =0 \Rightarrow \tl_r (a_k, b_k)=0, \quad k=1,2,
\end{equation}
that is, given a solution $\rho$ of the $\km_r$ equation (\ref{kmhie}), one obtains two
solutions, $(a_1, b_1)$ and $(a_2, b_2)$, of the $\tl_r$ equations (\ref{tlrabo}) related
to each other by the Miura-type transformations (\ref{defaot}), (\ref{defatt}).
For more information we refer to \cite{ghsz}, \cite{ttkm}, \cite{tjac}, and \cite{tw}.

\section{Equivalence of both constructions}
\label{secmain}

In this section we want to show that the constructions of the KM hierarchy outlined
in the previous two sections yield in fact the same set of evolution equations. This will
follow once we show that $G_j$ defined in (\ref{defGTL}) is the same as $G_j$ defined
in (\ref{defGKM}). It will be sufficient to consider the homogeneous quantities, however, we will
omit the additional hats for notational simplicity. Moreover, we will denote the sequence
$G_j$ defined in (\ref{defGTL}) by $\tilde{G}_j$ to distinguish it from the one defined in (\ref{defGKM}).
Since both are defined recursively via the recursions (\ref{reckmtl}) for $\tilde G_j, K_j$ respectively
(\ref{reckm}) for $G_j, H_j$ our first aim is to eliminate the additional sequences $K_j$ respectively
$H_j$ and to get a recursion for $\tilde G_j$ respectively $G_j$ alone.

\begin{lemma}
The coefficients $g_j(n)$ satisfy the following linear recursion
\begin{align} \label{4.2}
\begin{split}
g_{j + 3}^+ - g_{j + 3} &=
(b + 2 b^+) g_{j+2}^+ - (2 b + b^+) g_{j+2}\\ 
&\quad  {} - (2 b + b^+)b^+  g_{j+1}^+ + b (2 b^+ + b) g_{j+1} + k_{j+1}^+ + k_{j+1} \\ 
&\quad + b (b^+)^2 g_j^+- b^+ b^2 g_j- b k_j^+ - b^+ k_j,
\end{split}
\end{align}
where
\begin{equation}
k_j = a^2 g_j^+ - (a^-)^2 g_j^-, \quad j
\in\N.
\end{equation}
\end{lemma}

\begin{proof}
It suffices to consider the homogeneous case $g_j(n)=\spr{\delta_n}{H^j \delta_n}$.
Then (compare \cite[Sect~6.1]{tjac})
$$
g(z,n) = \spr{\delta_n}{(H-z)^{-1} \delta_n} = - \sum_{j=0}^\infty \frac{g_j(n)}{z^{j+1}}
$$
satisfies \cite[(1.109)]{tjac}
$$
\frac{(a^+)^2 g^{++} - a^2 g}{z-b^+} + \frac{a^2 g^+ - (a^-)^2 g^-}{z-b} =
(z-b^+) g^+ - (z-b) g,
$$
and the claim follows after comparing coefficients.
\end{proof}

\begin{corollary}
For $j\in \N_0$, the sequences $\tilde G_j$, defined by \eqref{defGTL} and corresponding to the $\tl$
hierarchy with $b\equiv0$, satisfy
\begin{align} \label{4.4}
\tilde G_{j+1}^+ -  \tilde G_{j+1} = (a^+)^2 \tilde G_j^{++} +
a^2(\tilde G_j^+ - \tilde G_{j}) - (a^-)^2\tilde G_{j}^-.
\end{align}
The corresponding sequences $G_j$ for the $\km$ hierarchy defined in \eqref{defGKM} satisfy
\begin{align} \label{4.5}
\begin{split}
G_{j+3} - G_{j+3}^{++}&= 
\big((a^-)^2 + a^2\big)^2 \big((a^+)^2 + (a^{++})^2\big) G_j \\
&\quad +(a^{--})^2 (a^-)^2 G_{j+1}^{--} + a^2 (a^+)^2 G_{j+1} \\
&\quad+\big((a^+)^2 + (a^{++})^2\big) \big(2(a^-)^2 + 2a^2 + (a^+)^2 + (a^{++})^2\big) G_{j+1}^{++}\\
&\quad+ \big(2(a^-)^2 + 2a^2 + (a^+)^2 + (a^{++})^2\big) G_{j+2}\\
&\quad- \big((a^-)^2 + a^2\big) \big((a^+)^2 + (a^{++})^2\big)^2 G_j^{++} \\
&\quad- \big((a^+)^2 + (a^{++})^2\big) \big((a^{--})^2 (a^-)^2 G_j^{--} - 
a^2 (a^+)^2 G_j^{++}\big)\\
&\quad - \big((a^-)^2 + a^2\big) \big(a^2 (a^+)^2 G_j - (a^{++})^2 (a^{+++})^2 G_j^{++++}\big) \\
&\quad- \big((a^-)^2 + a^2\big) \big((a^-)^2 + a^2 + 2(a^+)^2 + 2(a^{++})^2\big) G_{j+1}\\
&\quad - a^2 (a^+)^2 G_{j+1}^{++} - (a^{++})^2 (a^{+++})^2 
G_{j+1}^{++++}\\
&\quad- \big((a^-)^2 + a^2 + 2(a^+)^2 + 2(a^{++})^2\big) G_{j+2}^{++}.
\end{split}
\end{align}
\end{corollary}
\begin{proof}
Use  \eqref{4.2} with $b\equiv0$  for \eqref{4.4} resp.\ \eqref{defaot}, \eqref{defatt}
with $a=\rho$ for \eqref{4.5}.
\end{proof}

\begin{lemma}\label{l4.4}
For all $n\in\Z$,
\begin{equation}
\tilde G_j(n) = G_j(n), \quad j \in \N_0.
\end{equation}
\end{lemma}
\begin{proof}
Our aim is to show that $\tilde G_j$ satisfy the linear recursion relation \eqref{4.5} for $\hat G_j$. 
We start with \eqref{4.4},
\begin{align} \nn
\tilde G_{j+3} - \tilde G_{j+3}^+ + \tilde G_{j+3}^+ -  \tilde G_{j+3}^{++}
&= -(a^+)^2 \tilde G_{j+2}^{++} +
a^2(\tilde G_{j+2}-\tilde G_{j+2}^+) + (a^-)^2\tilde G_{j+2}^-\\ \label{4.7}
&\quad  -(a^{++})^2 \tilde G_{j+2}^{+++} +
(a^+)^2(\tilde G^+_{j+2}-\tilde G_{j+2}^{++}) + a^2\tilde G_{j+2},
\end{align}
and observe that the right hand side of \eqref{4.5} only involves even shifts of $G_j$.
Hence we systematically replace in \eqref{4.7} odd shifts of $\tilde G_j$ by \eqref{4.4}, 
\[
\tilde G_j=\left\{
\begin{array}{l}
G_{1,j} :=\tilde G_{j}^+ - (a^+)^2 \tilde G_{j-1}^{++} +
a^2(\tilde G_{j-1} - \tilde G_{j-1}^+) + (a^-)^2\tilde G_{j-1}^-\\
G_{2,j} := \tilde G_{j}^- + a^2 \tilde G_{j-1}^{+} +
(a^-)^2(\tilde G_{j-1} - \tilde G_{j-1}^-) - (a^{--})^2\tilde G_{j-1}^{--}
\end{array}\right.,
\]
as follows:
\[
\tilde G_{j+2}^{+++} \rightarrow G_{2, j+2}^{+++}, \quad 
\tilde G_{j+2}^{+} \rightarrow x G_{1, j+2}^{+} + (1-x)G_{2, j+2}^{+}, \quad
\tilde G_{j+2}^{-} \rightarrow G_{1, j+2}^{-},
\]
with
\[
x= \frac{(a^-)^2+a^2 + (a^{++})^2}{a^2-(a^+)^2}.
\]
In the resulting equation we replace
\[
\tilde G_{j+1}^{+++} \rightarrow G_{2, j+1}^{+++}, \quad 
\tilde G_{j+1}^{+} \rightarrow y G_{1, j+1}^{+} + (1-y)G_{2, j+1}^{+}, \quad
\tilde G_{j+1}^{-} \rightarrow G_{1, j+1}^{-},
\]
where
\[
y= \frac{(a^-)^2(a^{++})^2+a^2 (a^{++})^2}{a^2(a^{++})^2-(a^-)^2 (a^{+})^2}.
\]
This gives \eqref{4.5} for $\tilde G_j$.
\end{proof}

Hence both constructions for the KM hierarchy are equivalent and we have

\begin{theorem}
Let $r\in\N_0$. Then
\begin{equation}
\tl_{2r+1}(a,0)=\km_r(a). 
\end{equation}
provided $c^\tl_{2j+1}=c^\km_j$ and $c^\tl_{2j}=0$ for $j=0,\dots, r$.
\end{theorem}

\begin{remark}
As pointed out by M. Gekhtman to us, an alternate way of proving equivalence is
by showing that (in the semi-infinite case, $n\in\N$) both constructions give rise to the
same set of evolutions for the moments of the underlying spectral measure
(compare \cite{bs}). Our purely algebraic approach has the advantage that it does
neither require the semi-infinite case nor self-adjointness.
\end{remark}

\section{Appendix: Jacobi operators with $b \equiv 0$}
\label{app}

In order to get solutions for the Kac--van Moerbeke hierarchy out of solutions of the Toda hierarchy
one clearly needs to identify those cases which lead to Jacobi operators with $b \equiv 0$. For the
sake of completeness we recall some folklore results here.

Let $H$ be a Jacobi operator associated with the sequences $a$, $b$ as in (\ref{defjac}).
Recall that under the unitary operator $U f(n) = (-1)^n f(n)$ our Jacobi operator transforms
according to $U^{-1} H(a,b) U = H(-a,b)$, where we write $H(a,b)$ in order to display the
dependence of $H$ on the sequences $a$ and $b$. Hence, in the special case $b \equiv 0$
we infer that $H$ and $-H$ are unitarily equivalent, $U^{-1} H U = - H$. In particular,
the spectrum is symmetric with respect to the reflection $z \to -z$ and it is not surprising, that
this symmetry plays an important role.

Denote the diagonal and first off-diagonal of the Green's function of a Jacobi operator $H$ by
\begin{align}
\begin{split}
g(z,n) &=\spr{\delta_n}{(H-z)^{-1} \delta_n},\\
h(z,n) &=2a(n)\spr{\delta_{n+1}}{(H-z)^{-1} \delta_n} -1.
\end{split}
\end{align}
Then we have

\begin{theorem}
For a given Jacobi operator, $b \equiv 0$ is equivalent to $g(z,n)=-g(-z,n)$ and $h(z,n)=h(-z,n)$.
\end{theorem}

\begin{proof}
Set $\ti{H}= - U^{-1} H U$, then the corresponding diagonal and first off-diagonal elements
are related via $\ti{g}(z,n)=-g(-z,n)$ and $\ti{h}(z,n)=h(-z,n)$. Hence the claim follows since
$g(z,n)$ and $h(z,n)$ uniquely determine $H$ (see \cite[Sect.~2.7]{tjac} respectively \cite{ttr} for the
unbounded case).
\end{proof}

Note that one could alternatively use recursions: Since $g_j(n)$ and $h_j(n)$ are just the
coefficients in the asymptotic expansions of $g(z,n)$ respectively $h(z,n)$ around $z=\infty$
(see \cite[Chap.~6]{tjac}), our claim is equivalent to $g_{2j+1}(n)=0$ and $h_{2j}(n)=0$.

Similarly, $b \equiv 0$ is equivalent to $m_\pm(z,n)=-m_\pm(-z,n)$, where
\begin{equation}
m_\pm(z,n)=\spr{\delta_{n\pm 1}}{(H_{\pm,n} - z)^{-1} \delta_{n\pm 1}}
\end{equation}
are the Weyl $m$-functions. Here $H_{\pm,n}$ are the two half-line operators obtained from $H$
by imposing an additional Dirichlet boundary condition at $n$. The corresponding spectral measures
are of course symmetric in this case.

For a quasi-periodic algebro-geometric solution (see e.g.\ \cite[Chap.~9]{tjac}), this implies $b \equiv 0$ if and only if
both the spectrum and the Dirichlet divisor are symmetric with respect to the reflection $z \to - z$. For an
$N$ soliton solution this implies $b \equiv 0$ if and only if the eigenvalues come in pairs, $E$ and $-E$, and the norming
constants associated with each eigenvalue pair are equal.

\section*{Acknowledgments}

We thank Michael Gekhtman and Fritz Gesztesy for valuable discussions on this topic and hints
with respect to the literature.

\end{document}